%% file: DM.tex
\documentclass[11pt]{article}

 \usepackage{graphicx}
  \usepackage{epsfig}
  \usepackage{amssymb}
  \usepackage{subfigure}

\DeclareMathSymbol{\lesssim}      {\mathrel}{AMSa}{"2E}
\DeclareMathSymbol{\gtrsim}       {\mathrel}{AMSa}{"26}
\begin{document}
\def\be{\begin{equation}}
\def\ee{\end{equation}}
\def\bc{\begin{center}}
\def\ec{\end{center}}
\def\bea{\begin{eqnarray}}
\def\eea{\end{eqnarray}}
\def\dd{\displaystyle}
\def\nn{\nonumber}
\def\ad{\dot{\alpha}}
\def\ov{\overline}
\def\h4w{h_{4 w}}
\def\hlf{\frac{1}{2}}
\def\qrt{\frac{1}{4}}
\def\as{\alpha_s}
\def\at{\alpha_t}
\def\ab{\alpha_b}
\def\sq2{\sqrt{2}}
\newcommand{\smallz}{{\scriptscriptstyle Z}} %
\newcommand{\mz}{m_\smallz}
\newcommand{\smallw}{{\scriptscriptstyle W}}
\newcommand{\mw}{m_\smallw} 
\newcommand{\smallh}{{\scriptscriptstyle H}}
\newcommand{\mh}{m_\smallh}
\newcommand{\mt}{m_t}
\newcommand{\wh}{w_\smallh}
\newcommand{\wt}{w_t}
\newcommand{\zt}{z_t}
\newcommand{\toh}{t_\smallh}
\def\hto{h_t}
\def\zh{z_\smallh}
\newcommand{\Mvariable}[1]{#1}
\newcommand{\Mfunction}[1]{#1}
\newcommand{\Muserfunction}[1]{#1}
\thispagestyle{empty}
\begin{flushright}
RM3-TH/04-18 
\end{flushright}
\begin{center}
\vspace{1.7cm}
\bc
{\LARGE\bf Two-loop  electroweak corrections  to Higgs production 
at hadron colliders}

\ec
\vspace{1.4cm}
{\Large \sc    Giuseppe Degrassi$^{a}$ and Fabio Maltoni$^{b,a}$}

\vspace{1.2cm}

${}^a$
{\em 
Dipartimento di Fisica, Universit\`a di Roma Tre\\
INFN, Sezione di Roma III, Via della Vasca Navale~84, I-00146 Rome, Italy}
\vspace{.3cm}

${}^b$
{\em  Centro Studi e Ricerche ``Enrico Fermi'', via Panisperna 89/A, 
      I-00184 Rome,  Italy }

\end{center}

\vspace{0.8cm}

\centerline{\bf Abstract}
\vspace{2 mm}
\begin{quote}\small
We compute the two-loop electroweak corrections due to the top quark
to the gluon fusion production cross section of an intermediate-mass
Higgs boson. This result, together with the previously known
contribution due to the light fermions, allows a complete
determination of the two-loop electroweak corrections to this
production cross section. For Higgs mass values above 120 GeV the top
corrections have opposite sign of the light fermions contribution,
however they are always smaller in size reaching at most $15 \%$ of
the latter. The total electroweak corrections to the gluon fusion
production cross section for $115$ GeV $ \lesssim \mh \lesssim 2 \mw$
range from 5\% to 8\% of the lowest order term.

\end{quote}

\vfill
\newpage
\setcounter{equation}{0}
\setcounter{footnote}{0}
\vskip2truecm
\input{DM1.tex}

\input{DMbibl.tex}
\end{document}

%% file: DM1.tex
The Higgs bosons, the still missing particle of the Standard Model (SM)
responsible for the electroweak symmetry breaking, is actively searched for
at the Tevatron and it is one of the main objectives of experimental
program that  will be carried out at the future Large Hadron Collider (LHC)
which is supposed to span all the Higgs mass regions up to 1 TeV.  The
legacy of LEP has been a firm lower bound on the Higgs mass,
$\mh > 114$ GeV~\cite{unknown:2003ih}, and
at the same time, together with the information coming from SLD, a strong 
indirect evidence that the Higgs boson  should be relatively light with a high
probability for its mass to be below 200 GeV. A light Higgs is also
required by the supersymmetric extensions of the SM, the SUSY models,
which exhibit a definite prediction for the Higgs sector, namely that
the lightest CP-even SUSY Higgs particle should not be heavier than
140-150 GeV at most \cite{Allanach:2004rh}.

At hadron colliders the main Higgs production 
mechanism is the gluon fusion process \cite{H2gQCD0}.
Therefore, a precise knowledge of this process is very important 
in order to  put limits on the Higgs mass or, in case the Higgs is discovered,
to compare the measured cross section with the Standard Model (SM) result.
Given its relevance,  the gluon fusion production mechanism has received
in the recent years a large amount of theoretical work. Most of the
attention has been paid to the calculation of the QCD corrections.
The next-to-leading order (NLO) QCD corrections were first computed
in the heavy top limit \cite{H2gQCD1} and later keeping the full
top mass dependence  \cite{QCDg2}. Next-to-next-to-leading corrections
(NNLO) have also been computed, in the heavy top limit \cite{H2gQCD2}. 
A recent discussion \cite{bd4}
on the residual theoretical uncertainty from perturbative QCD contributions,
which includes also soft-gluon effects,
estimates it to be below 10\% for $\mh < 200 $ GeV.

Electroweak corrections to the gluon fusion 
production mechanism were first considered in the heavy top  limit
and found to give a very small effect, below 1\% \cite{DjG}. Recently,
the two-loop contribution due to the light fermions was  discussed
in Ref.\cite{ABDV}. That analysis shows that in the intermediate Higgs mass
region, i.e., $114 \: {\rm GeV} \lesssim \mh \lesssim 2\, \mw$, the 
contribution  of the  light fermions can be significant ranging from 
4\% to 9\% of the 
lowest order term. Instead, above the $2\, \mw$ threshold  the same analysis 
reports that the corrections are quite  small. 

Given the present status of the perturbative QCD uncertainty on the
gluon fusion production cross section, it seems relevant to have
a complete control of the electroweak corrections to this process at least
in the intermediate Higgs mass region. This requires, besides the 
knowledge of the light fermions contribution, the  result for
the two-loop top corrections\footnote{The contribution of the QED real 
corrections vanishes.}. The aim of this paper  is to present
such a contribution completing the calculation of the electroweak corrections
to the $gg \to H $ process. Our investigation applies to values of the Higgs
mass in the  intermediate  region. 

The Higgs boson has no tree-level coupling to
gluons; therefore the process $ g\,g \rightarrow  H$  is loop-induced.
At the partonic level the cross section, not corrected by QCD effects,
can be written as:
\bea
\sigma \left( g\,g \rightarrow  H  \right) & = &
\frac{G_\mu \alpha_s^2 }{512\, \sqrt{2} \, \pi}
\left| {\cal G} \right|^2 \, ,
\label{ggh}
\eea
where ${\cal G} = {\cal G}^{1l} + {\cal G}^{2l} + \dots$ and 
the lowest order one-loop  contribution is only due to the top quark and is
given by:
\bea
 {\cal G}^{1l} &=& -  \,4\,\toh \, \left[ 2 -  
(1 - 4\, \toh )\,
   H \left( -r,-r; -\frac1\toh \right) \right] ~,
\label{eq:onelooptop}
\eea
where $\toh \equiv \mt^2/\mh^2 $ 
and\footnote{All the analytic continuations are obtained with the replacement 
$x \rightarrow x -i\,\epsilon$}
\be
 H (-r,-r; x ) = \frac12
\log^2 \left( \frac{\sqrt{x+4}-\sqrt{x}}{\sqrt{x+4}+\sqrt{x}}
\right)~.
\label{eq:C0}
\ee

At the two-loop level the electroweak corrections to 
$\sigma \left( g\,g \rightarrow  H  \right) $ can be divided in two subsets: 
the corrections induced by the light (assumed massless) fermions and those 
due to the top quark, or
\be
 {\cal G}^{2l} =  {\cal G}^{2l}_{lf} +  {\cal G}^{2l}_t~.
\ee
The two contributions are very different because the former
one involves particles that  do not appear in the one-loop
calculation. In fact, at one loop, the light fermions do not contribute
to the gluon fusion process because of their small coupling to the Higgs boson.
Instead, at the two-loop level, the light fermions may couple 
to the $W$ or $Z$ bosons which in turn  can directly couple to the Higgs 
particle.
The light fermion contribution, $ {\cal G}^{2l}_{lf}$,
contains also diagrams where the bottom quark,
which is assumed massless, is present together with the $Z$ boson.
The light fermion corrections have been computed 
exactly in Ref.\cite{ABDV} where the analytic result has been expressed in 
terms of Generalized Harmonic Polylogarithms.

To compute the top contribution,
${\cal G}^{2l}_t$, we find it convenient to  employ the Background 
Field Method (BFM) quantization framework.
The BFM is a technique for quantizing gauge  theories \cite{BFM,abbott} that
avoids the complete explicit breaking of the gauge symmetry.
One of the salient features of this approach is that all fields are
splitted in two components: a classical background field $\hat{V}$
and a quantum field $V$ that appears only in the loops.
The gauge-fixing procedure is achieved through a non linear term in the
fields that breaks the gauge invariance only of the quantum part of the
lagrangian, preserving the gauge symmetry of the effective action with respect
to the background fields. Thus, in the BFM framework
some of the vertices in which  background fields are present are modified
with respect to the standard $R_\xi$ gauge ones. The  complete  set of
BFM Feynman rules for the SM can be found  in Ref.\cite{ddw}. 

The advantage of this quantization scheme is that in the BFM Feynman gauge 
(BFG), the one-particle irreducible
(1PI) contribution\footnote{The top mass counterterm diagrams are needed 
to cancel
divergencies in the two-loop irreducible corrections and, by definition, we 
include them in the 1PI contribution.} and the reducible one, i.e.\ the Higgs 
wave function renormalization plus the expansion of the bare coupling
$g_0/m_{\smallw_0}$ ($g$ being the $SU(2)$ coupling), are separately finite.
Concerning the latter, at the one-loop level $g_0/m_{\smallw_0}$ can be 
related to the $\mu$-decay constant via
\be
\frac{g_0}{m_{\smallw_0}} = \frac{8 \, G_\mu}{\sqrt{2}} \left( 1 -
        \frac12 \left[ - \frac{A_{\smallw \smallw}}{\mw} + V + B\, \right]
          \right)~~,
\ee
where $A_{\smallw \smallw}$ is the transverse part of the $W$ self-energy
at zero momentum transfer and the quantities $V$ and $B$ represent the
vertex and box corrections in the  $\mu$-decay amplitude. In the 
BFG  the combination 
\be
K_{r} =  \frac12 \left[  \frac{A_{\smallw \smallw}}{\mw} - V -B + 
                      \delta Z_\smallh\, \right] 
\label{Kew}
\ee
is finite where $\delta Z_\smallh$ is related to
the Higgs field wave function renormalization  through
\be 
H_0 = \sqrt{Z_\smallh} H \simeq 
 \left( 1 + \frac12\,\delta Z_\smallh \right) H~~.
\ee
We would like to stress that Eq.(\ref{Kew}) is finite only in the  
BFG  while in the standard $R_\xi$ Feynman gauge it 
shows some ultraviolet poles. 

Explicitly, we have in the  BFG and  in units $\alpha/(4 \pi s^2)$ 
\bea
K_r &=& N_c  \frac1{\wt}
\left[- \toh - \frac{1}{8}  +  ( \toh + \frac{1}{2}) 
\sqrt{4\toh-1} A(\toh) \right] 
     + \frac{13 - 2 \sqrt{3} \pi}{16 \wh}  \nn \\
  &&
     +\frac{(3 + 4 c^2) \log c^2 }{8 s^2}
     - \frac{3 \log\wh}{8 (1 - \wh)}
     +\frac{5 + 12 \zh}{16 c^2}
     +\frac{5 + 12 \wh}{8}   \nn \\
  && -3 \left(\frac{ \sqrt{4 \wh -1}}2 + \frac{ 2 \wh^2}{\sqrt{4 \wh -1}} 
          \right) A(\wh)
     -\frac3{2 c^2} \left( \frac{  \sqrt{4 \zh -1}}2 + 
     \frac{ 2 \zh^2}{ \sqrt{4 \zh -1}} \right) A(\zh), 
\label{Kr}
\eea
where $N_c$ is the color factor, $s^2 \equiv \sin^2 \theta_W$, 
$c^2 = 1- s^2,\, \wt \equiv  \mw^2 /\mt^2, \, \wh \equiv \mw^2 /\mh^2,$  
$\zh \equiv \mz^2 /\mh^2$ and 
\be
A(x) =  \arctan \frac{1}{\sqrt{4x-1}}~~.
\ee

The two-loop top contribution to the gluon fusion production cross section 
can be written as:
\be
 {\cal G}^{2l}_t =   K_{r}\, {\cal G}^{1l} +
                  {\cal G}^{2l}_{\rm 1PI}~~,
\label{G2t}
\ee 
where ${\cal G}^{2l}_{\rm 1PI}$ contains the the two-loop 1PI corrections.

\begin{figure}[t]
\begin{center}
\epsfxsize=10cm
\epsfbox{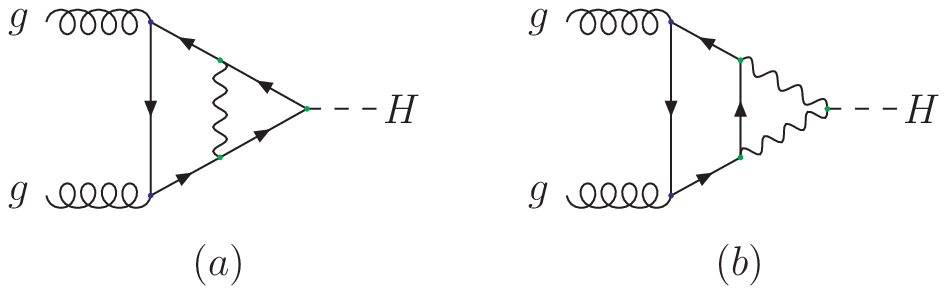}
\vspace*{0cm}
\end{center}
\caption{Examples of two-loop  diagrams contributing to $gg \to H$.}

\label{fig:one}
\end{figure}

To compute ${\cal G}^{2l}_{\rm 1PI}$  we notice that the diagrams contributing 
to it can be naturally organized in two classes:
i) diagrams with a triangular fermionic loop as well as top mass
counterterm diagrams,  that can be classified as corrections to the 
one-loop amplitude, like the one shown in Fig.(\ref{fig:one}a); ii) diagrams 
in which the Higgs does not couple directly to the top, 
Fig.(\ref{fig:one}b).
We notice that in the BFG the two sets of diagrams are 
separately finite and equal  to zero for vanishing Higgs mass.

\begin{figure}
\begin{center}
\epsfxsize=13cm
\epsfbox{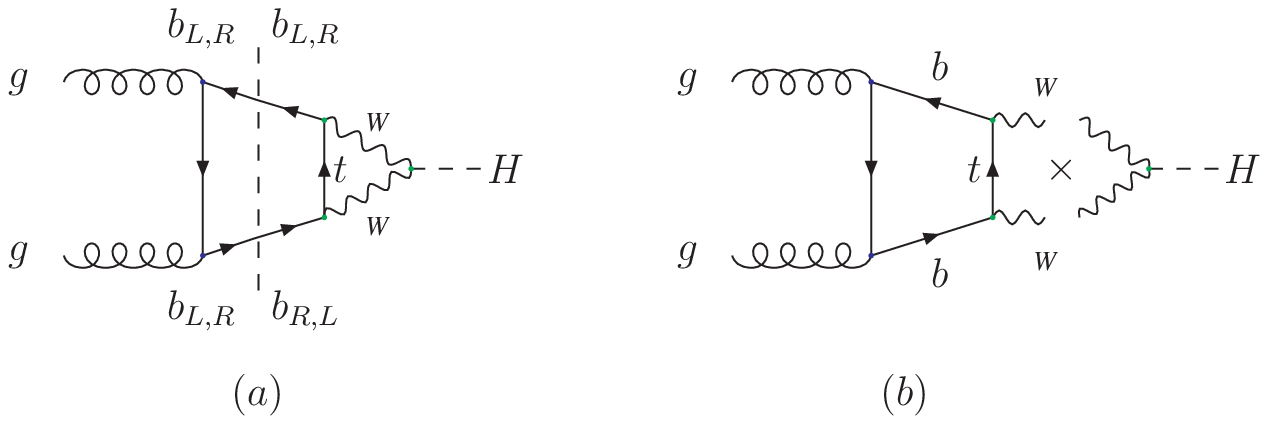}
\vspace*{0cm}
\end{center}
\caption{a) The possible intermediate helicities of the massless fermions
in a two-loop cut diagram. b) Amplitudes  corresponding to a cut 
at $q= 2 \mw$.}

\label{fig:two}
\end{figure}

To evaluate both kind of graphs we make the observation that, taken the 
bottom quark massless, some diagrams seem to have a cut at $q =0$, see 
Fig.(\ref{fig:two}a),  $q$ being the momentum carried by the Higgs. However,
this cut  is actually not present because of the helicity structure 
of the diagram. In fact,  the Higgs should  couple to one left-handed and 
one right-handed bottom quark, therefore the one-loop amplitude on the 
right-hand side of the dashed line in Fig.(\ref{fig:two}a) is non-zero only 
when the bottom quarks have opposite helicities, while in the tree amplitude on
the left-hand side the helicity is conserved along the quark line.
Since helicities cannot match, no cut develops at $q=0$.
In this situation, the first cut in 
these diagrams appears at $2 \, \mw$ (see Fig.(\ref{fig:two}b)).
Therefore, the evaluation of ${\cal G}^{2l}_{\rm 1PI}$ for Higgs mass in the 
intermediate region can be obtained by computing the relevant 
diagrams employing  an ordinary Taylor  expansion in the variable 
$\h4w \equiv  q^2/(4\, \mw^2)$.  

We follow this strategy to analytically evaluate the 1PI amplitudes up to 
$\h4w^4$ terms.  However, because the complete expression is quite long, we 
report explicitly only the leading term. We find in units 
$\alpha/(4 \pi s^2)$ 
\bea
{\cal G}^{2l}_{\rm 1PI} &=&
\frac{1}{\wt} \left[1 +
  \frac{ 2\toh +1}{4(4\toh -1)}
+\frac{\toh (2\toh -5) }{2(4\toh-1)^2} \log \toh 
-\frac{\toh (2\toh^2+1) }{2(4\toh-1)^2}\phi\left(\frac{1}{4 \toh}\right)
\right] -\frac{28}{9}-\frac{7}{18c^2} \nn\\
&&
+\frac{1}{\wh} \left[
  \frac{\wt -1-\log \wt}{3 (\wt-1)^2}
+\frac{5\zt -2}{12(\zt-4)^2}
+\frac{ 46 -19 \zt  }{6 (\zt-4)^3}\log \zt
+\frac{\zt^2-6 }{2\zt(\zt-4)^3}\phi\left(\frac{\zt}{4}\right)
\right] \nn\\
&&
+ {\cal O}(\h4w)~,
\eea
where $ \zt =  \mz^2 /\mt^2$ and 
\be
\phi(z) = 
       4 \sqrt{{z \over 1-z}} ~Cl_2 ( 2 \arcsin \sqrt z )~~,
       \label{e2.15c}
\ee
where $Cl_2(x)= {\rm Im} \,{\rm Li_2} (e^{ix})$ is the Clausen function.

\begin{table}
\label{tab1}
\addtolength{\arraycolsep}{0.1cm}
\renewcommand{\arraystretch}{1.3}
\begin{center}
\begin{tabular}[4]{c|c|cccc|c}
\hline
$\mh$ & topology & n=1 & 2 & 3 & 4 & ${\cal G}^{2l}_{\rm 1PI}$ \\
\hline
   & fer  & 1.4479 & 0.0762 & 0.0014 & -0.0001\\[-5pt]
\raisebox{1.5ex}[0cm][0cm]{115}& bos  & 0.8932 & 0.3552 & 0.0853 &  
0.0253& \raisebox{1.5ex}[0cm][0cm]{2.88}\\
\hline
   & fer  & 1.4887 & 0.0830 & 0.0017 & -0.0001\\[-5pt]
\raisebox{1.5ex}[0cm][0cm]{120}& bos  & 0.9662 & 0.4072 & 0.1079 &  
0.0348& \raisebox{1.5ex}[0cm][0cm]{3.08}\\
\hline
   & fer  & 1.5287 & 0.0900 & 0.0020 & -0.0001\\[-5pt]
\raisebox{1.5ex}[0cm][0cm]{125}& bos  & 1.0461 & 0.4648 & 0.1349 &  
0.0472& \raisebox{1.5ex}[0cm][0cm]{3.31}\\
\hline
   & fer  & 1.568  & 0.0974 & 0.0024 & -0.0002\\[-5pt]
\raisebox{1.5ex}[0cm][0cm]{130}& bos  & 1.1329 & 0.5287 & 0.1671 &  
0.0632& \raisebox{1.5ex}[0cm][0cm]{3.55}\\
\hline
   & fer  & 1.6064 & 0.1051 & 0.0028 & -0.0002\\[-5pt]
\raisebox{1.5ex}[0cm][0cm]{135}& bos  & 1.2265 & 0.5991 & 0.2052 &  
0.0836& \raisebox{1.5ex}[0cm][0cm]{3.82}\\
\hline
   & fer  & 1.6441 & 0.1131 & 0.0033 & -0.0003\\[-5pt]
\raisebox{1.5ex}[0cm][0cm]{140}& bos  & 1.3271 & 0.6767 & 0.2502 &  
0.1095& \raisebox{1.5ex}[0cm][0cm]{4.12}\\
\hline
   & fer  & 1.6811 & 0.1214 & 0.0038 & -0.0003\\[-5pt]
\raisebox{1.5ex}[0cm][0cm]{145}& bos  & 1.4345 & 0.7619 & 0.3029 &  
0.1420& \raisebox{1.5ex}[0cm][0cm]{4.44}\\
\hline
   & fer  & 1.7172 & 0.1301 & 0.0044 & -0.0004\\[-5pt]
\raisebox{1.5ex}[0cm][0cm]{150}& bos  & 1.5487 & 0.8552 & 0.3644 &  
0.1826& \raisebox{1.5ex}[0cm][0cm]{4.80}\\
\hline
   & fer  & 1.7526 & 0.1391 & 0.0050 & -0.0005\\[-5pt]
\raisebox{1.5ex}[0cm][0cm]{155}& bos  & 1.6699 & 0.9571 & 0.4358 &  
0.2329& \raisebox{1.5ex}[0cm][0cm]{5.19}\\
\hline
   & fer  & 1.7872 & 0.1484 & 0.0057 & -0.0006\\[-5pt]
\raisebox{1.5ex}[0cm][0cm]{160}& bos  & 1.7979 & 1.0683 & 0.5184 &  
0.2948& \raisebox{1.5ex}[0cm][0cm]{5.62}\\
\hline
\end{tabular}
\end{center}
\caption{Numerical values  of the $\h4w^{\rm n}$ terms in 
${\cal G}^{2l}_{\rm 1PI}$ in units  $\alpha/(4 \pi s^2)$ and their total. 
The label ``fer'' refers to diagrams of type i) while ``bos'' to 
type ii) ones. }
\end{table}

To appreciate the convergence of the series, we show
in table 1  the numerical values of the individual  terms in 
the expansion of ${\cal G}^{2l}_{\rm 1PI}$. We have further separated them
in the two contributions corresponding to the diagrams of type i), labeled
``fer'',  and ii) labeled ``bos''. The  input values chosen in the table 
are:  $\mt= 178$ GeV, $\mw = 80.4$ GeV and $\mz= 91.18$ GeV. 
As expected, the ``fer'' contribution  shows a very good 
convergence for Higgs masses  in the intermediate region.  
In fact,  in these diagrams the first cut actually occurs at $q = 2 \mt$. 
In the same Higgs mass region, the ``bos''part has a worse convergence. 
However, taking the last term
in the expansion as error of the calculation we find that for 
$\mh = 160$ GeV we can assign to the ``bos'' contribution  an uncertainty of
$9 \%$, which  translates into a $5 \%$ uncertainty on  the total top 
correction.  Clearly, in the case of lower Higgs mass the 
uncertainty will be reduced. 

\begin{table}[t]
\label{tab2}
\addtolength{\arraycolsep}{0.1cm}
\renewcommand{\arraystretch}{1.4}
\begin{center}
\begin{tabular}[4]{c|cc|c}
\hline
$\mh$ & 1 light gen & $3^{\rm rd}$ gen & $\delta_{ew} (\%)$ \\
\hline
115   & -5.28 & -0.78 - 0.22 & 4.7\\
120   & -5.62 & -0.82 - 0.06 & 4.9\\
125   & -5.98 & -0.87 + 0.12 & 5.1\\
130   & -6.36 & -0.93 + 0.33 & 5.4\\
135   & -6.76 & -0.98 + 0.58 & 5.6\\
140   & -7.20 & -1.04 + 0.88 & 5.8\\
145   & -7.69 & -1.10 + 1.26 & 6.1\\
150   & -8.26 & -1.16 + 1.78 & 6.4\\
155   & -9.01 & -1.23 + 2.68 & 6.6\\
160   & -10.4 & -1.30 + 3.43 & 7.5\\
\hline
\end{tabular}
\end{center}
\caption{Contributions to the amplitude, in units  $\alpha/(4 \pi s^2)$, 
of one massless (from Ref.\cite{ABDV}), and of the third
generation and total relative correction to the cross section. The
third generation contribution has been divided into a topless part
(from Ref.\cite{ABDV}, left number) and the rest (Eq.(\ref{G2t}),
right number). }
\end{table}

Our result on the top corrections can be put together with the result of 
Ref.\cite{ABDV} on the light fermion contribution to obtain a 
complete prediction of the two-loop electroweak correction to the
gluon fusion process for Higgs mass values in the intermediate region.
In table 2, we compare the numerical values of the two-loop corrections
to the amplitude due to a light fermion generation  with that of the
third generation. Concerning the latter, we have divided it
in two parts. The first number corresponds to  diagrams not containing
the top, which have been computed exactly in Ref.\cite{ABDV}. The second one
instead corresponds to $ {\cal G}^{2l}_t$, as calculated in this paper.  
The table shows that the contribution of the third generation is at most 
$20 \%$ of that of a light fermion. We notice that, in the units used in 
the table, the result of the large $\mt$ limit \cite{DjG} amounts to 
$-\mt^2/(6 \mw^2) = -0.82$. This number  does not approximate at all the total
correction, which is actually dominated by the light fermion contribution, and
it is also quite different from the result of $ {\cal G}^{2l}_t$ even in the 
case  of low Higgs mass. Concerning  $ {\cal G}^{2l}_t$, we notice that the
two terms in Eq.(\ref{G2t}) tend to cancel each other, cf. table 1 and 2.
In table 2 we also report the result for
$\mh =160$ GeV, namely in  the  mass region close to the $2\, \mw$ 
threshold that actually requires a more refined analysis. In fact,
close to the threshold the factor $K_r$ in the top corrections 
behaves as $1/\sqrt{1- 4\,\wh}$ for $\wh$ close to $1/4$, cf. Eq.(\ref{Kr}). 
This unphysical singularity is related to the opening 
of the $2 \, W$ channel in the wave-function renormalization of the Higgs 
\cite{Kniehl:1993ay} and it is a 
signal that in this region a first order treatment of the $W$-boson propagator 
is inadequate. A way to eliminate this threshold singularity is to employ the
definition of the mass and of the width of a particle from the complex-valued
position of its propagator's pole \cite{KPS}. Following this procedure
we  replace $\mw^2$ with $\mw^2 - i \Gamma_\smallw \mw$ in the 
$A(\wh)/\sqrt{4 \wh -1}$ term appearing in Eq.(\ref{Kr}).
In the same table  we have also reported the  total electroweak relative 
corrections $\delta_{ew}$ to the Higgs production cross  section 
$\sigma \equiv \sigma_0 (1+\delta_{ew})$, where $\sigma_0$ is the lowest order 
result. 

The result we have derived for $\sigma \left( g\,g \rightarrow  H  \right)$
can be easily translated into  a result for the corrections to
the partial decay width $\Gamma ( H \rightarrow g g)$ recalling the
relation
\be
\Gamma \left( H \rightarrow  g g  \right)  = 
\frac{8 \mh^3}{\pi^2} \, \sigma(g g\rightarrow H) \, .
\label{h2g}
\ee

In conclusion, we have computed the top contribution to 
$\sigma(g g\rightarrow H)$ for an intermediate-mass Higgs particle completing
the calculation of the two-loop electroweak 
correction to this important production cross section. In this Higgs mass 
range the electroweak corrections 
are dominated by the light fermion contribution computed in Ref.\cite{ABDV}. 
The top contribution has opposite sign to it, apart from the Higgs mass region 
close to the present experimental lower bound,  but in any case it is much 
smaller in size reaching at most $15 \%$ of the light fermion contribution. 

To complete the calculation of the electroweak corrections to 
$\sigma(g g\rightarrow H)$ for any value of the Higgs mass the top corrections
for $\mh > 2 \mw$ are still needed. However, for this  Higgs mass values
the light fermion corrections are quite small and we do not
expect  the top corrections to have a size  significantly different 
from the $\mh < 2 \mw$ case. Thus, the total electroweak correction
to $\sigma(g g\rightarrow H)$ for  $\mh > 2 \mw$ is probably quite
small.

The authors want to thank U.~Aglietti for interesting discussions.
This work was partially supported by the European Community's
Human Potential Programme under contract
HPRN-CT-2000-00149 (Physics at Colliders).